\documentclass[11pt,a4paper]{article}
\usepackage{pdfsync}
\usepackage{subcaption}
\usepackage{booktabs}

\usepackage{natbib}
\usepackage{amsfonts,amsmath,amsthm}
\usepackage{bm}

\numberwithin{theorem}{section}

\theoremstyle{definition}

\usepackage{amssymb}
\usepackage{hyperref}
\usepackage{graphicx}
\usepackage{wrapfig}
\usepackage{epsfig}
\usepackage{float,mathrsfs}
\usepackage{enumerate}
\usepackage{enumitem}
\usepackage{bbm}
\usepackage{comment}
\usepackage[normalem]{ulem}
\newtheorem{algorithm}{Algorithm}
\usepackage{algorithm}
\usepackage{algpseudocode}
\usepackage[font=scriptsize]{caption}
\counterwithout{figure}{section}
\usepackage{color}
\definecolor{refkey}{rgb}{0.9451,0.2706,0.4941}\definecolor{labelkey}{rgb}{0.9451,0.2706,0.4941}

\definecolor{darkred}{RGB}{139,0,0}
\definecolor{darkgreen}{RGB}{0,100,0}
\definecolor{darkmagenta}{RGB}{139,0,139}
\definecolor{gray}{RGB}{180,180,180}

\newcommand{\mask}[1]{{}}

\usepackage{pgfplots}
\pgfplotsset{compat=1.18}

\title{Efficient Amortized Bayesian Inference for Markov Random Fields via Gradient-Informed Grid Selection}

\author{L. Bazahica\footnotemark[1]~\footnotemark[2] \and A. Avalos-Pacheco\footnotemark[3]~\footnotemark[4] \and M. T. Moores\footnotemark[2]\and L. Roininen\footnotemark[2]}

\allowdisplaybreaks

\begin{document}
\footnotetext[1]{corresponding author}
\footnotetext[2]{School of Engineering Sciences, LUT University, P.O.~Box 20, 53851 Lappeenranta, Finland ({\tt laura.bazahica@lut.fi, lassi.roininen@lut.fi, matthew.moores@lut.fi}).}
\footnotetext[3]{Institute of Applied Statistics, JKU Linz, Altenberger Straße 69, 4040 Linz, Austria ({\tt alejandra.avalos\_pacheco@jku.at})}
\footnotetext[4]{Harvard-MIT Center for Regulatory Science, Harvard University, 200 Longwood Ave, Boston, MA 02115, USA}

\maketitle

\begin{abstract}
    \noindent Bayesian inference for models with intractable likelihoods, such as Markov random fields, poses a fundamental computational challenge due to the tradeoff between inferential accuracy and computational cost. Various MCMC methods have been developed to address this challenge. The exchange algorithm targets the exact posterior, but requires an expensive perfect sampling step at each iteration, which is often infeasible in practice. In contrast, path sampling approximates the Metropolis acceptance ratio using a precomputed grid of likelihood values, but may introduce bias when the grid is poorly chosen. 
    We introduce a novel amortized MCMC framework that retains the theoretical validity of exact methods while substantially reducing the computational burden. The proposed approach employs a gradient–informed grid selection procedure and constructs a surrogate likelihood via Hermite interpolation, yielding a smooth approximation with low error. A simulation study characterizes the rate at which inferential accuracy improves as the number of grid points increases. We further demonstrate the practical performance of the method through applications to a hidden Potts model for satellite imagery and an autologistic model for Arctic ice floes.
\end{abstract}
\vspace{2pc}

\section{Introduction}

Markov random field (MRF) models are widely used to represent complex dependency structures across a range of applications, including image analysis \citep{winkler2012image}, survey data \citep{avalos2025bayesian,Avalos2026ABMI}, epidemiology \citep{besag1991bayesian},  and social networks \citep{kolaczyk2014statistical}. 
Despite their flexibility, Bayesian inference for many MRFs is severely hindered by the presence of intractable normalizing constants in the likelihood. These constants depend on the model parameters and cannot be evaluated analytically, rendering the likelihood unavailable in closed form and the resulting posterior distribution doubly intractable.
Well-known examples include the Potts model \citep{potts1952some} and the exponential random graph model  \citep[ERGM;][]{frank1986markov}. This intractability poses a fundamental obstacle to statistical inference, since most algorithms require the likelihood to be evaluated pointwise at each iteration.\\

Many approaches 
have been developed to address this challenge \citep{moores2020bayesian,park2018bayesian}. For example, the exchange algorithm is a Markov Chain Monte Carlo (MCMC) method that uses an auxiliary Markov chain for forward simulation from the intractable model at each iteration \citep{murray2012mcmc}. While the exchange algorithm can produce unbiased estimates for certain models, it relies on perfect sampling \citep{propp1996exact,huber2016perfect}, which has a high computational cost and is often infeasible for practical applications. The approximate exchange algorithm \citep[AEA;][]{cucala2009bayesian} mitigates the need for perfect sampling by substituting it with Gibbs sampling \citep{geman1984stochastic} or with the algorithm of \citet{swendsen1987nonuniversal}. While this approximation reduces computational burden, it introduces some bias by only running the auxiliary chain for a fixed number of iterations. Moreover, despite this simplification, the computational cost of AEA remains high, making it impractical for many real-world applications.\\

A faster alternative is the path sampling algorithm, also known as thermodynamic integration (TI). \citet{gelman1998simulating} derived an approximation to the log-ratio of the model's normalizing constants, enabling the precomputation of sufficient statistics that can be interpolated rather than recomputed at each MCMC iteration. The accuracy of this approach depends critically on the quality of the interpolation, and a variety of surrogate models have been proposed to address this issue. \citet{calderhead2009estimating} showed that Bayes factors can be estimated accurately and efficiently by combining TI with population MCMC, even for complex, high-dimensional models. The controlled thermodynamic integral introduced by \citet{oates2016controlled} further reduces the variance of Bayesian model evidence estimates, improving the stability and efficiency of marginal likelihood computation. \\

The parametric functional approximate Bayesian (PFAB) algorithm of \cite{moores2020scalable} was developed specifically for the hidden Potts model, achieving posteriors estimates comparable in accuracy to the exchange algorithm while substantially reducing runtime. However, its design does not generalize easily to models with higher-dimensional parameter spaces. A recent extension by \cite{vu2025warped} employs warped gradient-enhanced Gaussian process surrogate models to handle multidimensional parameters, but relies on the assumption that the sufficient statistics of the exponential family model are independent, an assumption that is often violated in practice. In a different line of work, \cite{boland2018gibbs} proposed new estimators for the Metropolis acceptance ratio based on a precomputed grid of sufficient statistics values. While this approach has a solid theoretical foundation for applications in higher dimensions, the grid size grows exponentially with the number of parameters, making the precomputation step computationally prohibitive.\\

Identifying an optimal set of grid points in a computationally efficient manner is a major challenge. In Bayesian optimization \citep[BO;][]{jones1998efficient}, points are selected sequentially to optimize an acquisition function using surrogate models, typically based on Gaussian processes. For high-dimensional parameter spaces, approaches such as variable selection and low-dimensional embedding exploit problem structures to make BO feasible \citep{wang2023recent}. Despite these advances, the placement of grid points remains a difficult high-dimensional optimization problem.\\

In this work, we introduce a new framework for efficiently precomputing and interpolating a grid of sufficient statistics. At each MCMC iteration, the interpolated values are used to compute the Metropolis acceptance ratio, providing a faster alternative to both simulating the sufficient statistics and evaluating the path sampling identity. Unlike previous methods, our approach leverages gradient-informed grid selection and Hermite interpolation to achieve high accuracy across the parameter space. This results in posterior estimates comparable to those of the exchange algorithm, while substantially reducing computational cost. In doing so, our method effectively combines the strengths of both path sampling and approximate exchange algorithms, offering a scalable and generalizable solution for Bayesian inference with doubly-intractable likelihoods.\\

This paper is organized as follows. We define the theoretical setting for our models in section \ref{sec:model}. In section \ref{sec:methodology} we introduce MCMC methods for Bayesian inference in Markov Random fields and continue with algorithms for precomputation in section \ref{sec:precomp}. A simulation study as well as applications to a hidden Potts model of satellite data and the autologistic model of Arctic ice floes are showcased in section \ref{sec:numexp}. In section \ref{sec:conclusion} conclusions and future prospects are presented.

\section{Markov random field models}\label{sec:model}

In this paper we focus on MRF models where the likelihood can be expressed as an exponential family, e.g.
\begin{align}\label{eq:expFamily}
    p(\bm z\mid \bm\beta)=\frac{\exp{(\bm\beta^TS(\bm z))}}{\mathcal{C}(\bm\beta)},
\end{align}
where $\bm z=(z_1, \dots, z_N)$ are the given data, $\bm \beta=(\beta_1, \dots, \beta_D)^T$ are the model parameters, $S(\bm z)=(s_1(\bm z), \dots, s_D(\bm z))^T$ are the sufficient statistics and the normalizing constant
\begin{align}\label{eq:normConstant}
    \mathcal{C}(\bm \beta)=\sum_{\bm z \in \mathcal{Z}}\exp{(\bm\beta^TS(\bm z))}
\end{align} is the sum over all possible configurations of discrete states $\bm z \in \mathcal{Z}$. Evaluating this constant becomes infeasible for large $\mathcal{Z}$. There exist many models of this form including the hidden Potts model and the autologistic model that we discuss in the following. 

\subsection{Hidden Potts model}\label{sec:hiddenpotts}
Let $\bm y \in \mathbb{R}^n$ be the observed pixels of an image. The goal is to label theses pixels by a finite set of discrete states $\bm{z} \in \mathcal{Z}:=\{1, \dots, k\}^n$. 
In the Potts model the labels follow a Gibbs distribution that is specified by the conditional probabilities
\begin{align}\label{eq:zi}
    p(z_i\mid z_{\setminus i}, \bm\beta) = \frac{\exp{(\bm\beta\,\sum_{l\in \partial(i)}\delta(z_i, z_l))}}{\sum_{j=1}^k\exp{(\bm\beta\,\sum_{l\in \partial(i)}\delta(j, z_l))}}, \quad i\in \{1, \dots n\}
\end{align}
where $\bm \beta$ is called the inverse temperature, $z_{\setminus i}$ represents all labels except $z_i$, $\partial (i)$ are the neighboring pixels of $i$, and $\delta(u, v)$ is the Kronecker delta function. The sum in the upper exponential term is therefore a count of the neighbors of pixel $i$ with the same label, while the sum in the lower term counts and sums the exponential of the number of neighbors of $i$ that are labeled state $j$ for all states $j\in \{1, \dots, k\}$. \\

\begin{figure}
\centering
\begin{subfigure}{0.48\textwidth}
    \includegraphics[width=\textwidth]{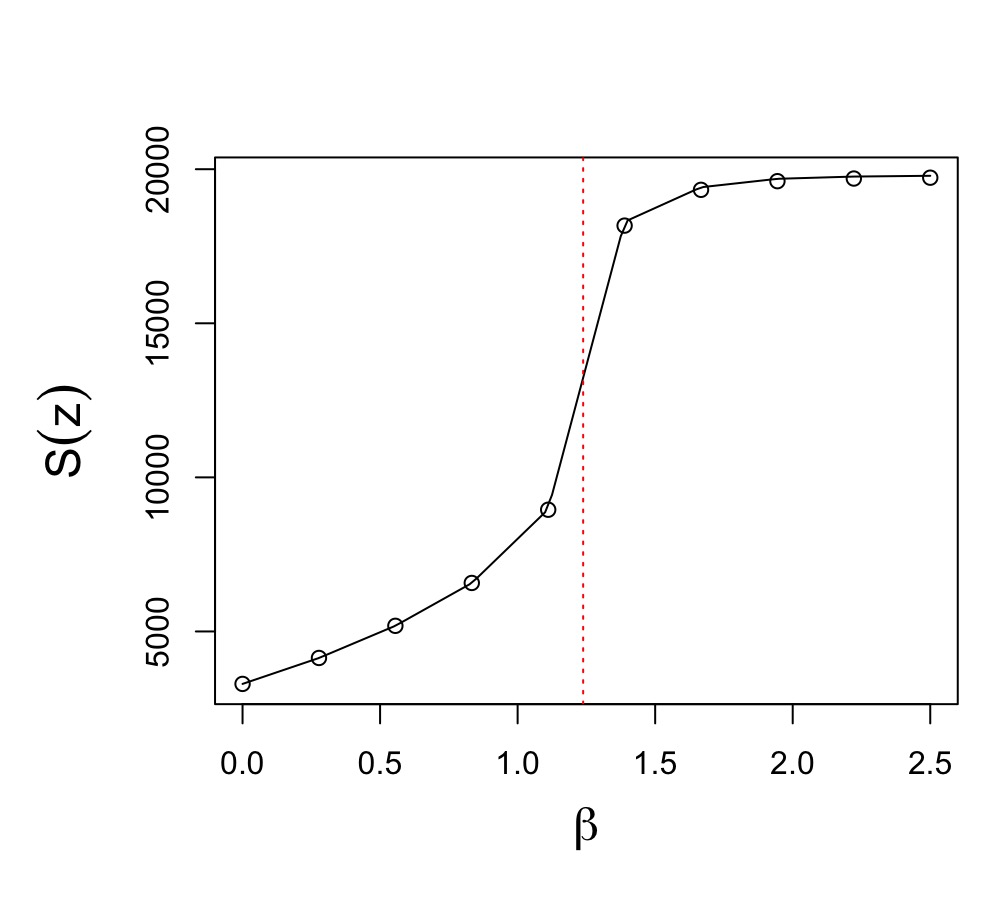}
    \caption{Linear interpolation of equidistant grid}
    \label{fig:LinLin}
\end{subfigure}
\hfill
\begin{subfigure}{0.48\textwidth}
    \includegraphics[width=\textwidth]{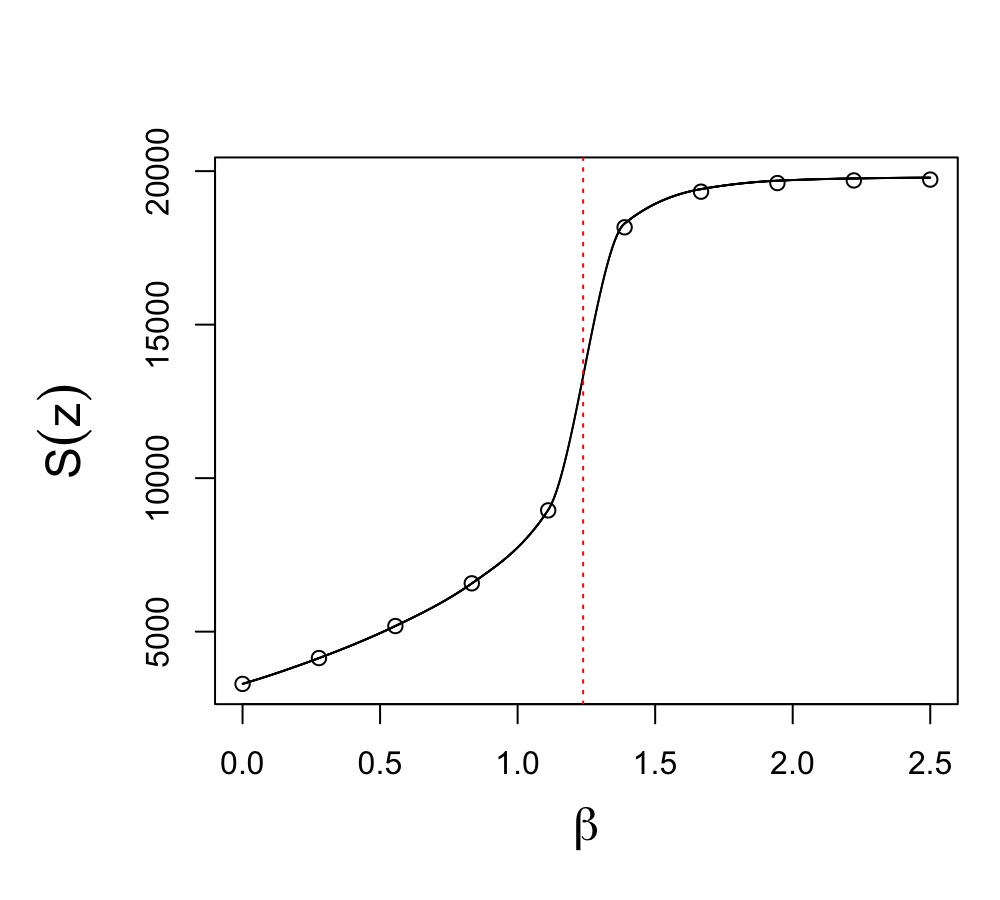}
    \caption{Hermite interpolation of equidistant grid}
    \label{fig:LinHer}
\end{subfigure}
\hfill
\begin{subfigure}{0.48\textwidth}
    \includegraphics[width=\textwidth]{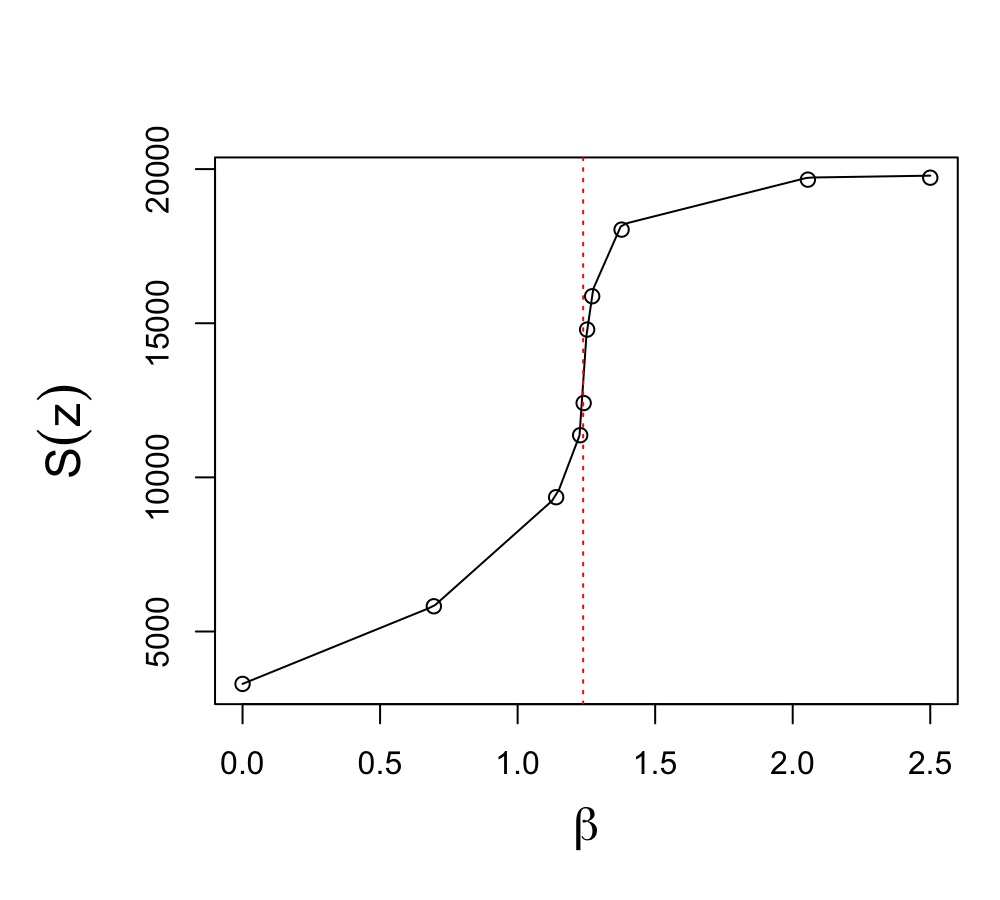}
    \caption{Linear interpolation of gradient-based grid}
    \label{fig:ExpLin}
\end{subfigure}
\hfill
\begin{subfigure}{0.48\textwidth}
    \includegraphics[width=\textwidth]{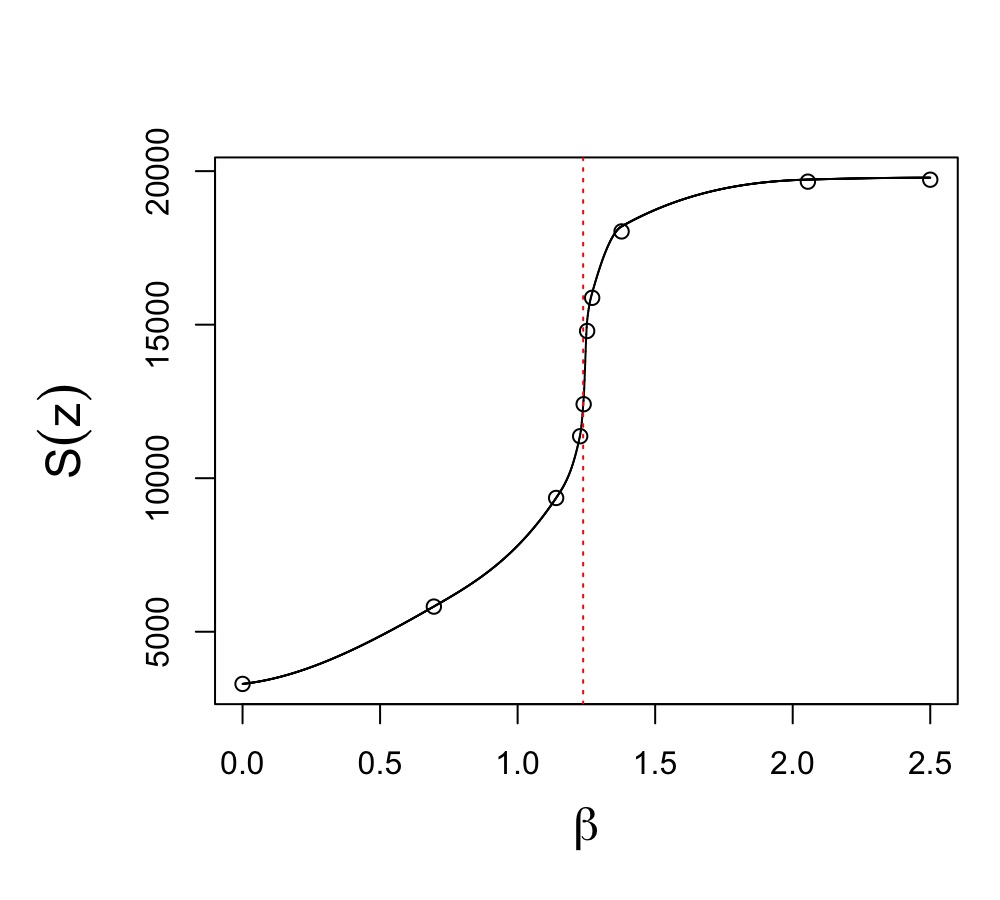}
    \caption{Hermite interpolation of gradient-based grid}
    \label{fig:ExpHer}
\end{subfigure}
        
\caption{Expected values of $S(\bm z)$, all estimated at 10 points in $[0,2.5]$ for $\bm\beta$ in a hidden Potts model with $k=6$}
\label{fig:figures}
\end{figure}

Since the labels $\bm{z}$ of the pixels are not directly observed in the hidden Potts model, the parameters $\theta_{z_i}$ govern the distribution of the pixel values with the label $z_i$. Then the observation equation for the pixel values is given by
\begin{align}\label{eq:obs}
    p(\bm y \mid\bm z, \bm \theta) =\prod_{i=1}^n p(y_i \mid z_i, \theta_{z_i}),
\end{align}
where $\bm y$ is the observed pixel value and $n$ is the number of pixels. It holds, that $\theta_{z_i} = (\mu_{z_i}, \sigma_{z_i}^2)$, such that
\begin{align}
    y_i\mid z_i= j, \mu_j, \sigma_j^2 \, \sim \mathcal{N}(\mu_j, \sigma_j^2),
\end{align} 
where $\{\mu_j\}_j, \{\sigma_j^2\}_j$ are unknown and equipped with informative prior distributions chosen based on the available data as described in \citet{moores2020scalable}.
The joint distributions of the labels is of the form of \eqref{eq:expFamily} with $S(\bm z)=\sum_{\langle i,l\rangle \in \mathcal E}\delta(z_i, z_l)$ where $\mathcal{E}$ is the set of neighboring pixels and $\mathcal{C}(\bm\beta)$ as defined in \eqref{eq:normConstant}. The joint posterior in this setting is 
\begin{align}
    p(\bm \theta, \bm \beta, \bm z \mid\bm y) \propto p(\bm y\mid\bm z, \bm \theta)\,p(\bm \theta)\,p(\bm z\mid\bm \beta)\,p(\bm \beta),
\end{align}
where $p(\bm\theta)$ is the joint prior for the parameters in \eqref{eq:obs} and $p(\bm \beta)$ is the prior for the inverse temperature. Since it is feasible to sample from \eqref{eq:zi}, one can estimate the expected value of $S(\bm z)$ for a fixed $\bm \beta$ as illustrated in Figure \ref{fig:figures} by employing the algorithm proposed by \citet{swendsen1987nonuniversal}.

\subsection{Autologistic model}\label{sec:autologistic}

Another well-known instance of an MRF is the autologistic model. Autologistic models are commonly applied to binary data exhibiting auto-correlated responses. We focus on the two-state autologistic model introduced by \cite{besag1974spatial}, whose likelihood is given by
\begin{align}\label{eq:autologistic}
    p(\bm z\mid\bm \beta)= \frac{\exp{\bigl(\beta_{1}\sum_u z_u + \beta_{2}\sum_{u\sim v}\delta(z_u, z_v)\bigr)}}{\mathcal{C}(\bm \beta)},
\end{align}
where $z_u\in \{-1,1\}$ represents the label of pixel $u$ and $\bm \beta = (\beta_1,\beta_2)^T$ is the two-dimensional set of model parameters. In addition to counting neighboring pixels that share the same label, the model also encodes the total number of pixels assigned to each label. Typical applications include modeling the spatial distribution of wildlife \citep{augustin1996autologistic} and characterizing spatial dependence in satellite imagery of ice floes \citep{vu2025warped}.

\section{MCMC methods and algorithms}\label{sec:methodology}

MCMC methods are a common approach to do Bayesian inference for many models \citep{brooks2011handbook}. However, for models with intractable distributions as introduced in Section \ref{sec:model} the calculation of posterior distributions becomes far more challenging. Even for this setting a growing number of feasible algorithms exist \citep{park2018bayesian}. \\

In the following we introduce a version of the exchange algorithm, which we will use as quality control in our numerical experiments, and path sampling, whose main idea to use simulated values in the acceptance ratio is the baseline for our approach.

\subsection{Approximate exchange algorithm}\label{sec:exchange}

The exchange algorithm proposed by \cite{murray2012mcmc} as an improvement of the work of \cite{moller2006efficient} avoids the need for a fixed estimate for the normalizing constant $\mathcal{C}(\bm\beta)$ by introducing an auxiliary variable so that $\mathcal{C}(\bm\beta)$ cancels out in the Metropolis-Hastings (M-H) ratio. This requires perfect sampling from the stationary distribution of the Potts model, which is possible \citep{huber2016perfect}, but increases the runtime of the algorithm significantly. Algorithm \ref{alg:ex} details the method proposed by \cite{cucala2009bayesian}. The function $q(\cdot \mid \cdot)$ denotes the M-H proposal density.

\begin{algorithm}
\caption{Approximate Exchange Algorithm (AEA)}\label{alg:ex}
    \begin{algorithmic}[1]
        \State Initialize $\beta_0$, $\bm z$
        \For{$t \in 1, \dots, T$}
        \State Update the label $z_i\sim p(z_i\mid z_{\setminus i}, \beta) \quad \forall i \in \{1, \dots, n\}$
        \State Compute sufficient statistics $S(\bm z)$
        \State Draw proposed noise parameter value $\beta'\sim q(\beta'\mid\beta_{t-1})$
        \State Generate $\bm w\mid\beta'$ by sampling from \eqref{eq:expFamily}
        \State Calculate the M-H acceptance ratio 
        \begin{align}
            \rho =\min\bigg(1, \frac{q(\beta_{t-1}\mid\beta')\,\exp(\beta'^TS(\bm{z}))\,\mathcal{C}(\beta_{t-1})}{q(\beta'\mid\beta_{t-1})\, \exp(\beta_{t-1}^TS(\bm z))\,\mathcal{C}(\beta')}\frac{\exp(\beta_{t-1}^TS(\bm{w}))\, \mathcal{C}(\beta')}{\exp(\beta'^TS(\bm{w}))\,\mathcal{C}(\beta_{t-1})}\bigg)
        \end{align}
        \State Draw $u\sim$Uniform$[0,1]$
        \If{$u<\rho$}
        \State $\beta_t \leftarrow \beta'$
        \Else{ $\beta_t\leftarrow \beta_{t-1}$}
        \EndIf
        \EndFor
    \end{algorithmic}
\end{algorithm}

\subsection{Path sampling / Thermodynamic integration}

Path sampling or thermodynamic integration (TI) has proven to be more scalable than the exchange algorithm for large datasets, while still obtaining comparable results. This is due to the additional computations of $S(\bm z)$ being done in a precomputation step rather than in every iteration of the MCMC algorithm. A throwback is the dependence on the choice of suitable $\bm\beta$-values to interpolate the $\bm\beta \rightarrow \mathbb{E}_{\bm z \mid\bm\beta}[S(\bm z)]$ function.

The path sampling identity presented in \cite{gelman1998simulating},
\begin{align}\label{eq:pathID}
    \log\biggl\{\frac{\mathcal{C}(\beta_{t-1})}{\mathcal{C}(\beta')}\biggr\} = \int_{\beta'}^{\beta_{t-1}}\mathbb{E}_{\bm z \mid\bm\beta}[S(\bm z)]\,\text{d}\bm\beta
\end{align}
is approximated by generating samples from \eqref{eq:expFamily} at a predetermined grid of $\bm\beta$ values. The resulting estimates of $\mathbb{E}_{\bm z \mid\bm\beta}[S(\bm z)]$ are then used to numerically approximate the integral in \eqref{eq:pathID}, and this approximation is subsequently substituted into the $\log$ M–H ratio in Algorithm \ref{alg:path}. Figure \ref{fig:LinLin} shows a piecewise-linear interpolation of $\mathbb{E}_{\bm z \mid\bm\beta}[S(\bm z)]$ on a two-dimensional lattice for $k = 6$ labels, with $\bm \beta$ varying from 0 to 2.5 in steps of $0.2\overline{7}$. The approximation was precomputed using the method of \citet{swendsen1987nonuniversal}.

\begin{algorithm}
\caption{Path sampling (thermodynamic integration)}
\label{alg:path}
\begin{algorithmic}[1]
    \State Initialize $\beta_0$, $\bm z$
    \For{$t = 1, \dots, T$}
        \State Update the variables $z_i \sim p(z_i \mid z_{\setminus i}, \beta_{t-1}), \quad \forall i \in \{1, \dots, n\}$
        \State Compute sufficient statistics $S(\bm z)$
        \State Draw proposed parameter value $\beta' \sim q(\beta' \mid \beta_{t-1})$
        \State Estimate $\mathbb{E}_{\bm z\mid\bm\beta}[S(\bm z)]$ for $\bm\beta \in \{\beta', \beta_{t-1}\}$ by interpolation
        \State Approximate the path sampling integral from \eqref{eq:pathID}
        \State Compute the log Metropolis--Hastings acceptance ratio:
        \begin{align}
            \log \rho
            =
            \min \Bigg(0, \log\biggl\{\frac{\mathcal{C}(\beta_{t-1})}{\mathcal{C}(\beta')}\biggr\}
            + (\beta' - \beta_{t-1})^\top S(\bm z) +
            \log\biggl\{\frac{q(\beta_{t-1} \mid \beta')}
                 {q(\beta' \mid \beta_{t-1})}
            \biggr\}
            \Bigg)
        \end{align}
        \State Draw $u \sim \text{Uniform}(0,1)$
        \If{$u < \rho$}
            \State $\beta_t \leftarrow \beta'$
        \Else{ $\beta_t \leftarrow \beta_{t-1}$}
        \EndIf
    \EndFor
\end{algorithmic}
\end{algorithm}

\section{Precomputation and interpolation}\label{sec:precomp}

The work of \citet{boland2018gibbs} has shown that it is possible to accurately estimate the posterior by using precomputed values in the MCMC ratio even without calculating the path sampling identity. The introduced algorithm computes the grid points around the mode of the posterior and uses information of gradient and hessian to choose optimal directions. The number of directions equals two times the parameter dimension. It is not clear if the algorithm returns different grids depending on the choice of starting direction. Additional constant parameters are a control for step size and a stopping criterion. A too strong stopping criterion means that proposed parameter values in the MCMC chain for obtaining posterior estimates are not necessarily covered by the grid, making it impossible to calculate the acceptance ratio. While a small step size is a good fit close to the posterior mode, to guarantee full coverage of the parameter space without an exploding grid size, we propose an adaptive control, which will also allow for a weaker stopping criterion.

\subsection{Grid choices}\label{sec:pathchoices}

We investigate and compare how efficient different constructions of the grid $\mathcal{G} = \{\beta^1, \dots, \beta^l\}$ are in the precomputation step. As a first, straightforward choice, we consider an equidistant grid over the parameter space $\mathcal{P}$. This approach has a long history in the literature of path sampling \citep{marin2007bayesian, moores2015external}. While calculation of grid points is straight forward and grid size is easily controllable by the practitioner, grid size scales exponentially with dimension and critical minimal number of needed grid points to obtain accurate posterior estimates is not clear.\\ 

Our second strategy, based on gradient information of the sufficient statistics $S(\bm z)$, starts building a grid from a point $\beta_{crit}$, obtained either from model-specific prior knowledge or from an estimate of the posterior mode computed via a stochastic approximation algorithm \citep{robbins1951stochastic}. 
Using a model-informed starting point decouples the grid from the specific data to be evaluated and allows it to be reused for additional data sets in $\mathcal{Z}$. Following \citet{boland2018gibbs}, we then approximate the eigenvectors of the inverse Hessian of the log-posterior at this starting point. These eigenvectors provide a basis of directions $d_1, \dots, d_D$ in $\mathcal{P}$ used in the grid selection algorithm \ref{alg:grad_path}.\\ 

\begin{algorithm}
\caption{Gradient Grid Choice}\label{alg:grad_path}
    \begin{algorithmic}[1]
        \State Initialize $\beta_{crit}$, $\mathcal{G} =\{\beta_{crit}\}$
        \State Estimate $\mathbb{E}_{\bm z\mid\beta_{crit}}[S(\bm z)]$, $\nabla_{\beta_{crit}}$, $n_{\beta_{crit}}$ and $d_1, \dots, d_D$
        \For{$s=1, \dots, D$}
        \State Initialize $\hat{\mathcal{G}} =\emptyset$
        \For{$\hat{\bm \beta}\in \mathcal{G}$}
        \State Initialize $n_{\hat{\bm \beta}}, \,\bm \beta = \hat{\bm \beta}-n_{\hat{\bm \beta}}d_s$
        \While{$\bm \beta\in \mathcal{P}$}
        \State Set $\hat{\mathcal{G}} = \hat{\mathcal{G}}\cup\bm\beta$
        \State Estimate $\mathbb{E}_{\bm z\mid\bm\beta}[S(\bm z)]$ and $\nabla_{\bm \beta}$
        \State Calculate $n_{\bm\beta}$
        \State Set $\bm \beta = \bm\beta-n_{\bm\beta}d_s$
        \EndWhile
        \State Initialize $n_{\hat{\bm \beta}}, \,\bm\beta = \hat{\bm \beta}+n_{\hat{\bm \beta}}d_s$
        \State Repeat steps 7-12 with $\bm\beta=\bm \beta+n_{\bm \beta}d_s$ instead of $\bm\beta = \bm\beta-n_{\bm\beta}d_s$
        \EndFor
        \State Set $\mathcal{G} = \mathcal{G}\cup \hat{\mathcal{G}}$
        \EndFor
        \State Return $\mathcal{G}$
    \end{algorithmic}
\end{algorithm}

We determine the function $n_{\bm \beta}$ for scaling the directional vectors to the next $\bm\beta$-value using a model-based rule that depends on the gradient $\nabla_{\bm \beta}$ of the expected values of the sufficient statistics $S(\bm z)$ at current $\bm \beta$. Our proposal for the hidden Potts model is
\begin{align}\label{eq:expstep}
    n_{\bm\beta} =  \exp{(-\kappa\,\nabla_{\bm\beta}/\nabla_{\beta_{crit}})}, \quad \kappa >0,
\end{align}
motivated by expressions proposed in \cite{white2016predicting} for applications in environmental epidemiology and supported by earlier work such as \cite{diggle1997regression}, where it outperformed all competing methods. In models with one dimensional parameters, this quantity directly corresponds to the step size between grid points. The scaling function $n_{\bm\beta}$ is chosen to be strictly non-negative, monotonically decreasing, and finite at points $\bm\beta$ with $\nabla_{\bm\beta}=0$, which are all desirable properties. The parameter $\kappa$ not only determines the step size but also provides a handle on the overall number of grid points.\\

\begin{figure}
    \centering
    \begin{subfigure}{0.49\textwidth}
        \includegraphics[width=\textwidth]{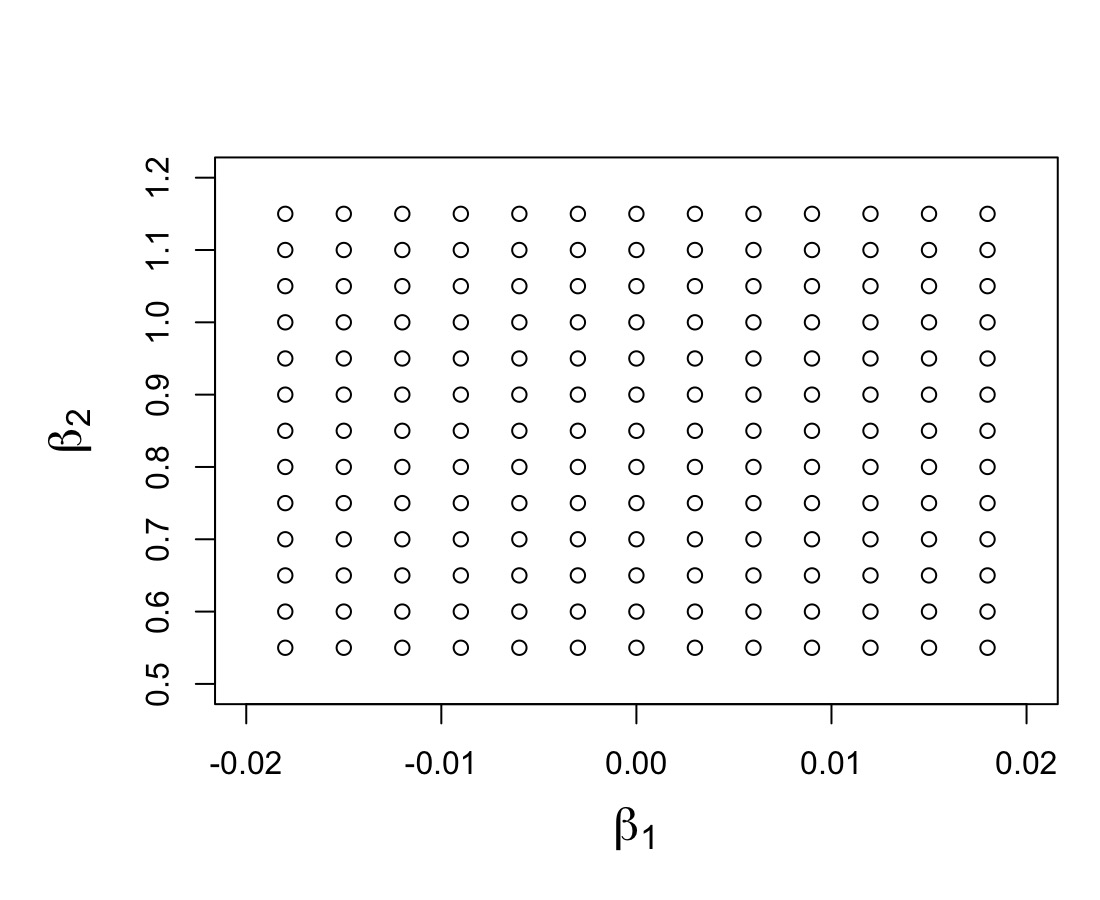}
        \caption{Equidistant grid with 169 points}
    \end{subfigure}
    \begin{subfigure}{0.49\textwidth}
        \includegraphics[width=\textwidth]{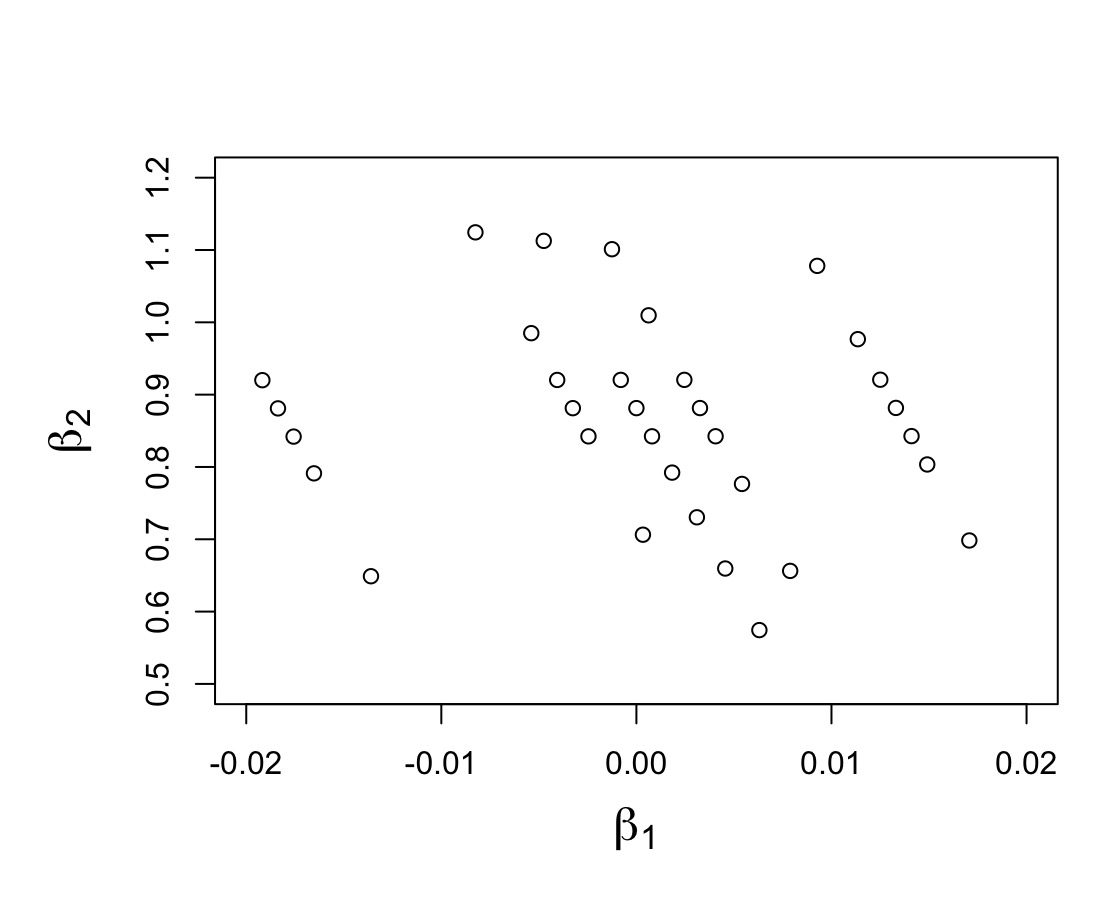}
        \caption{Gradient-based grid with 33 points}
    \end{subfigure}
    \caption{Precomputed grids for the autologistic example}
    \label{fig:2dgrid}
\end{figure}

We stop adding grid points in a particular direction once the corresponding boundary of the parameter space $\mathcal{P}$ is reached. An equidistant grid can be recovered by algorithm \ref{alg:grad_path} by choosing $D$-dimensional unit vectors as directional vectors and a constant scaling function $n_{\bm \beta}$. A comparison of equidistant and gradient-based grids for a two-dimensional parameter in an autologistic model is shown in Figure \ref{fig:2dgrid}.

\subsection{Interpolation of the sufficient statistics}

Interpolation methods offer advantages over regression methods in that they require relatively little computation while still achieving high accuracy, and they also come with theoretical error bounds \citep{lux2021interpolation}. We examine two possible approaches for interpolating the recovered grid points. The first approach relies on piecewise-linear interpolation, which has already been employed by \citet{marin2007bayesian} to evaluate the path sampling identity, yielding satisfactory results. The second approach is based on Hermite coordinate interpolation \citep{kechriniotis2024classical}. Although this method is somewhat more computationally demanding than piecewise-linear interpolation, it produces smoother solutions, particularly in non-linear settings. This improvement stems from the use of derivative information at the interpolation points. A comparison of piecewise-linear versus Hermite interpolation on equidistant and gradient-based grids in a hidden Potts model with $k=6$ labels is shown in Figure \ref{fig:figures}.

\begin{table}[ht!]
\centering
\begin{tabular}{|l|l|l|l|l|l|l|}
\hline
\textbf{Grid type}      & \textbf{Interpolation} & \textbf{5 pts} & \textbf{10 pts} & \textbf{20 pts} & \textbf{40 pts} & \textbf{60 pts}\\ \hline
\textit{equidistant}    & \textit{linear}        & 1569              & 398        & 207        & 94   &   62           \\ \hline
\textit{equidistant}    & \textit{hermite}            & 1405              & 278     & 138    & 63   &     49         \\ \hline
\textit{gradient-based} & \textit{linear}         & 1031              & 348        & 306       & 252  &   154         \\ \hline
\textit{gradient-based} & \textit{hermite}            & 866              & 82          & 116         & 62     &        34    \\ \hline
\end{tabular}
\caption{RMSE of for different grid sizes}
\label{tab:msenewpath}
\end{table}

\section{Experiments}\label{sec:numexp}

We explore the accuracy of our proposed grids and interpolation methods for obtaining estimates of $\mathbb{E}_{\bm z\mid\bm\beta}[S(\bm z)]$ in a simulation study and through the resulting posterior estimates in a hidden Potts model and an autologistic model. We compare to instances of the AEA, the Gaussian surrogate model introduced by \citet{vu2025warped}, classical path sampling and one of the algorithms proposed in \citet{boland2018gibbs} demonstrating that our method achieves comparable results with substantially lower computational cost. To compare the different posterior estimates we calculate posterior mean and modes as well as the Kullback-Leibler (KL) divergence \citep{kullback1951information}. Estimates of the sufficient statistics of the models are obtained using the Swendsen-Wang algorithm for one-dimensional parameters and Gibbs sampling for two-dimensional parameters. In the figures and tables we refer to the combinations of grid and interpolation methods in the following way: EL for equidistant grid and piecewise-linear interpolation, EH for equidistant grid and Hermite interpolation, GL for gradient-based grid and piecewise-linear interpolation and GH for gradient-based grid and Hermite interpolation. All analyses were conducted in \texttt{R} 4.3.1 on Ubuntu 24.04.

\subsection{Simulation study}

Regardless of the specific method, increasing the size of the precomputed grid improves the quality of the interpolations and thus the overall results. We simulate grids of various sizes to demonstrate that our procedures yield useful grids starting from relatively small point sets, which in turn reduces computation time, and that poor reconstructions of the sufficient statistic over the full parameter space are negligible for accurate posterior estimations as long as the critical region around $\beta_{crit}$ is adequately covered.\\

For the one-dimensional parameter in the hidden Potts model with $k=6$ labels we generate a $100\times100$ image with true $\bm \beta = 1.274$. After adding noise to the image, we estimate $\mathbb{E}_{\bm z \mid \bm \beta}[S(\bm z)]$ for different values of $\bm\beta$ chosen according to our methodology in section \ref{sec:precomp}. We take points from the parameter space $\mathcal{P} = [0,2.5]$ with varying grid sizes $|\mathcal{G|} \in \{5,10,20,40,60\}$ which corresponds to the directional vector $d_1=1$ and the constant scaling function $n_{\bm \beta}\in \{\frac{2.5}{4}, \frac{2.5}{9}, \frac{2.5}{19}, \frac{2.5}{39}, \frac{2.5}{59}\}$ in the equidistant approach. In the gradient-based approach we propose the critical value $\beta_{crit}=\log{(1+\sqrt{k})}$ derived by \citet{potts1952some} as a suitable choice for the starting point. Through variation of $\kappa$ in \eqref{eq:expstep} we recover 5 grids with sizes $|\mathcal{G|} \in \{5,10,20,40,60\}$.\\ 

\begin{figure}
    \centering
    \includegraphics[width=0.5\linewidth]{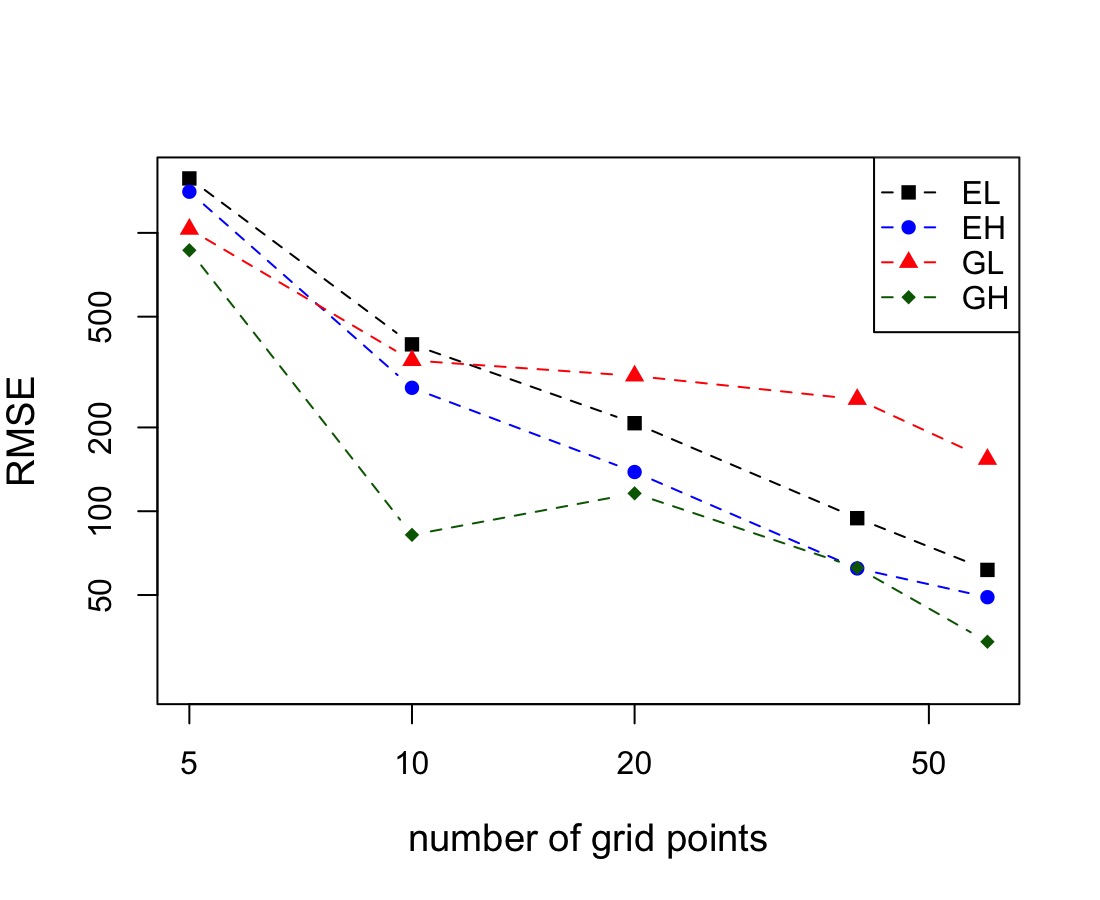}
    \caption{loglog-plot of RMSE for different grid sizes in a hidden Potts model}
    \label{fig:rmse1d}
\end{figure}

To evaluate the interpolation methods, we use the estimates of the sufficient statistics at the grid points to interpolate $\mathbb{E}_{\bm z \mid \bm \beta}[S(\bm z)]$ over $\mathcal{P}$. We then compare the interpolated values to direct estimates of $\mathbb{E}_{\bm z \mid \bm \beta}[S(\bm z)]$ computed at a test set of 1,000 randomly selected points $\bm \beta \in \mathcal{P}$. Specifically, we compute residuals between the interpolated values obtained from the precomputed grids and the corresponding direct estimates at these points. The results in Table \ref{tab:msenewpath} indicate that, for small grid sizes, Hermite interpolation more accurately recovers the expected values of $S(\bm z)$ than piecewise-linear interpolation, as measured by the root mean squared error (RMSE). The results are displayed in Figure \ref{fig:rmse1d}, which also shows that the advantages of a gradient-based grid for recovering $\mathbb{E}_{\bm z\mid\bm\beta}[S(\bm z)]$ only hold in combination with Hermite interpolation.\\

Additionally, we obtain posterior estimates of $\bm\beta$ for all grid sizes, types and interpolation methods and compare their mean and KL divergence to a run of the AEA in Table \ref{tab:postMeanMode}. Computation time was $7.04$ hours for the AEA and $90$ seconds per posterior for the precomputational approaches. In a slight contradiction to the RMSE results, here both gradient-based approaches return desired and better results than equidistant grids, which supports our strategy to focus grid points on critical regions of $\mathbb{E}_{\bm z\mid\bm\beta}[S(\bm z)]$. 

\begin{table}[ht]
\centering
\begin{tabular}{lcccccccccc}
\toprule
& \multicolumn{2}{c}{5} 
& \multicolumn{2}{c}{10} 
& \multicolumn{2}{c}{20} 
& \multicolumn{2}{c}{40} 
& \multicolumn{2}{c}{60} \\
\cmidrule(lr){2-3} 
\cmidrule(lr){4-5}
\cmidrule(lr){6-7}
\cmidrule(lr){8-9}
\cmidrule(lr){10-11}
& Mean & KL 
& Mean & KL 
& Mean & KL 
& Mean & KL 
& Mean & KL \\
\midrule
EL & 1.446 & 31.486 & 1.326 & 31.486 & 1.295 & 24.220 & 1.277 & 0.726 & 1.272 & 0.158 \\ 
  EH & 1.396 & 31.486 & 1.305 & 29.828 & 1.286 & 6.517 & 1.274 & 0.238 & 1.271 & 0.437 \\ 
  GL & 1.440 & 31.486 & 1.274 & 0.014 & 1.276 & 0.048 & 1.275 & 0.035 & 1.275 & 0.076 \\ 
  GH & 1.407 & 31.486 & 1.271 & 0.184 & 1.273 & 0.023 & 1.272 & 0.094 & 1.273 & 0.041 \\ 
\bottomrule
\end{tabular}
\caption{D mean and KL divergence to a run of the AEA for different grid sizes, grid types and interpolation types, where $31.486$ is the upper bound due to our specific implementation}
\label{tab:postMeanMode}
\end{table}

\subsection{Hidden Potts example}

\begin{figure}
    \centering
    \begin{subfigure}{0.45\textwidth}
        \includegraphics[width=\textwidth]{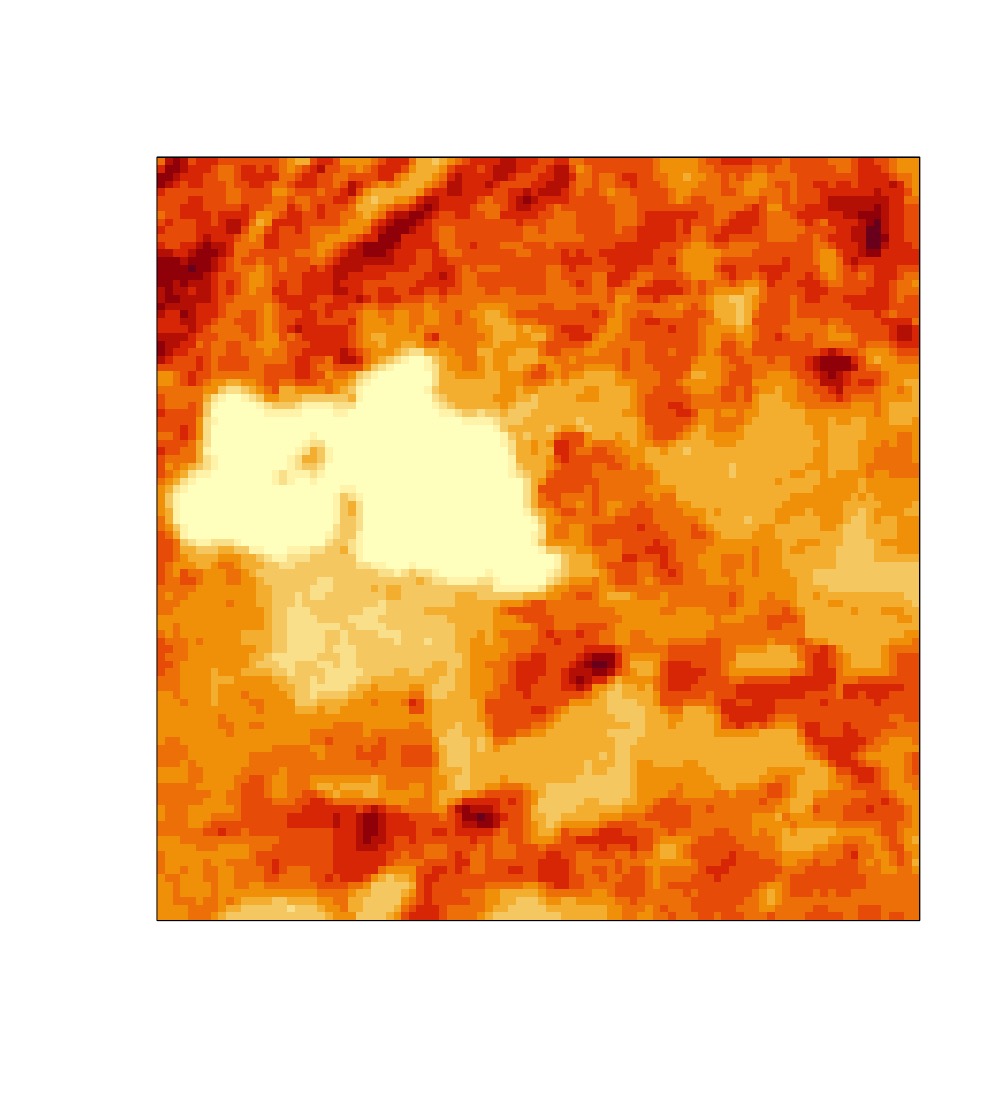}
        \vspace{-1cm}
        \caption{Lake Menteith data}
        \label{fig:MenteithData}
    \end{subfigure}
    \begin{subfigure}{0.47\textwidth}
        \includegraphics[width=\textwidth]{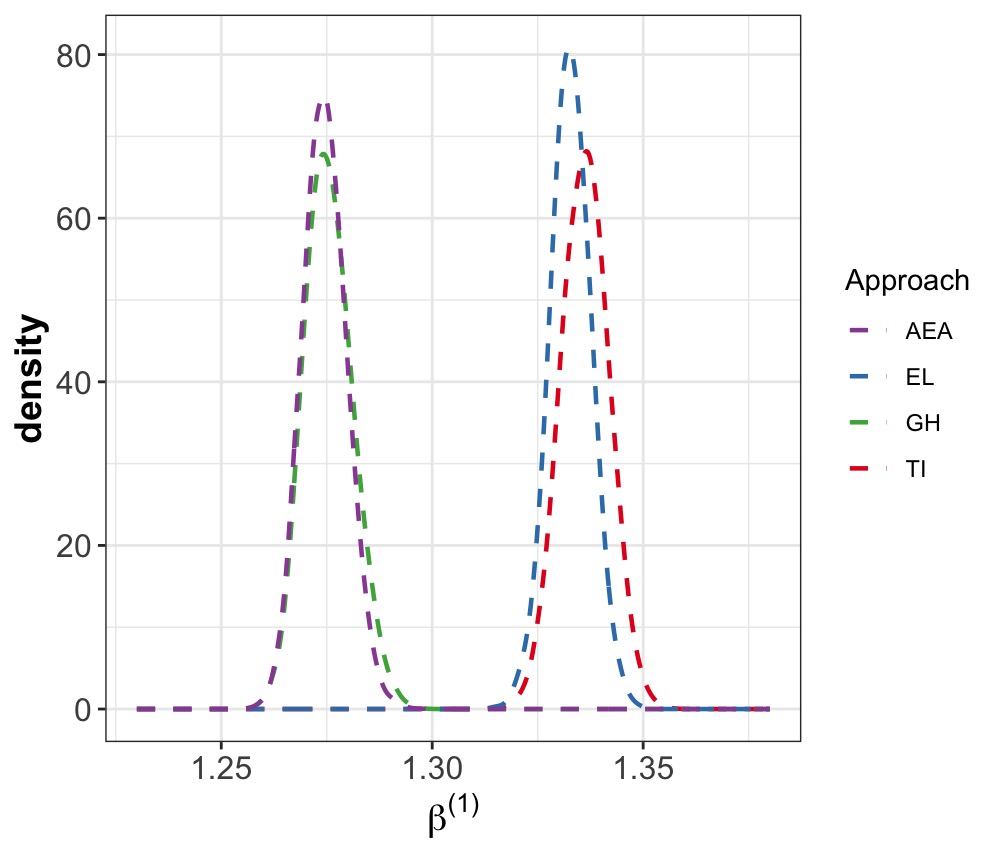}
        \caption{Posterior estimates for the Menteith example}
        \label{fig:PostMenteith}
    \end{subfigure}
    \caption{Lake Menteith example}
    \label{fig:PostBeta}
\end{figure}

A well known example for path sampling in the hidden Potts model is the classification of the pixels of a $100\times100$ satellite image of Lake Menteith in Scotland into $k=6$ labels. We use the dataset \texttt{Menteith} available in the \texttt{R} package \texttt{Bayess} \citep{BayessR} to compare our methods. The dataset is displayed in Figure \ref{fig:MenteithData}. As a gold standard for the posterior we take a sufficiently long run of the AEA. 
We precompute equidistant and gradient-based grids with 10 points each with a computation time of 6 seconds. 
In Figure \ref{fig:PostMenteith} we display the estimated posterior distributions obtained through the AEA, classical path sampling with $10$ $\beta$-values, which we will refer to as TI in the following, the EL and the GH approach. Computation times excluding precomputation were 6.17 hours for the AEA, 0.02 hours for TI, 0.03 hours for GH and 0.03 hours for EL.
The KL divergence to the exchange algorithm was 31.38 for the TI approach, 31.52 for the equidistant grid-based approach and 0.03 for the gradient-based grid approach.

\subsection{Autologistic example}

\begin{figure}
    \centering
    \begin{subfigure}{0.35\textwidth}
        \includegraphics[width=\textwidth]{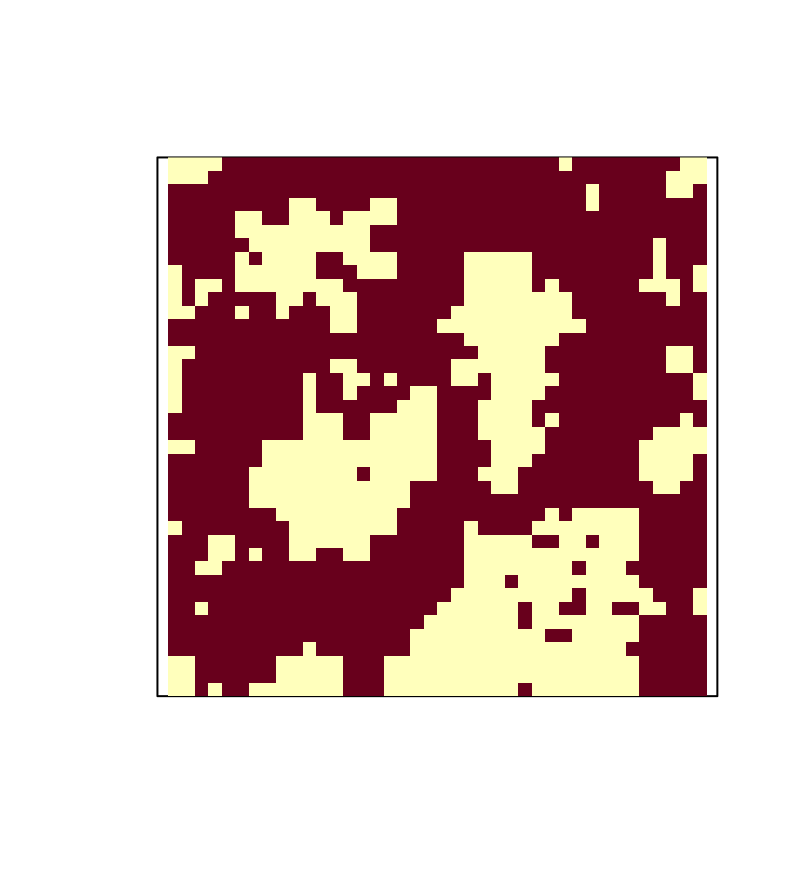}
        \caption{Ice floe data}
        \label{fig:IceFloeData}
    \end{subfigure}
    \begin{subfigure}{0.63\textwidth}
        \includegraphics[width=\textwidth]{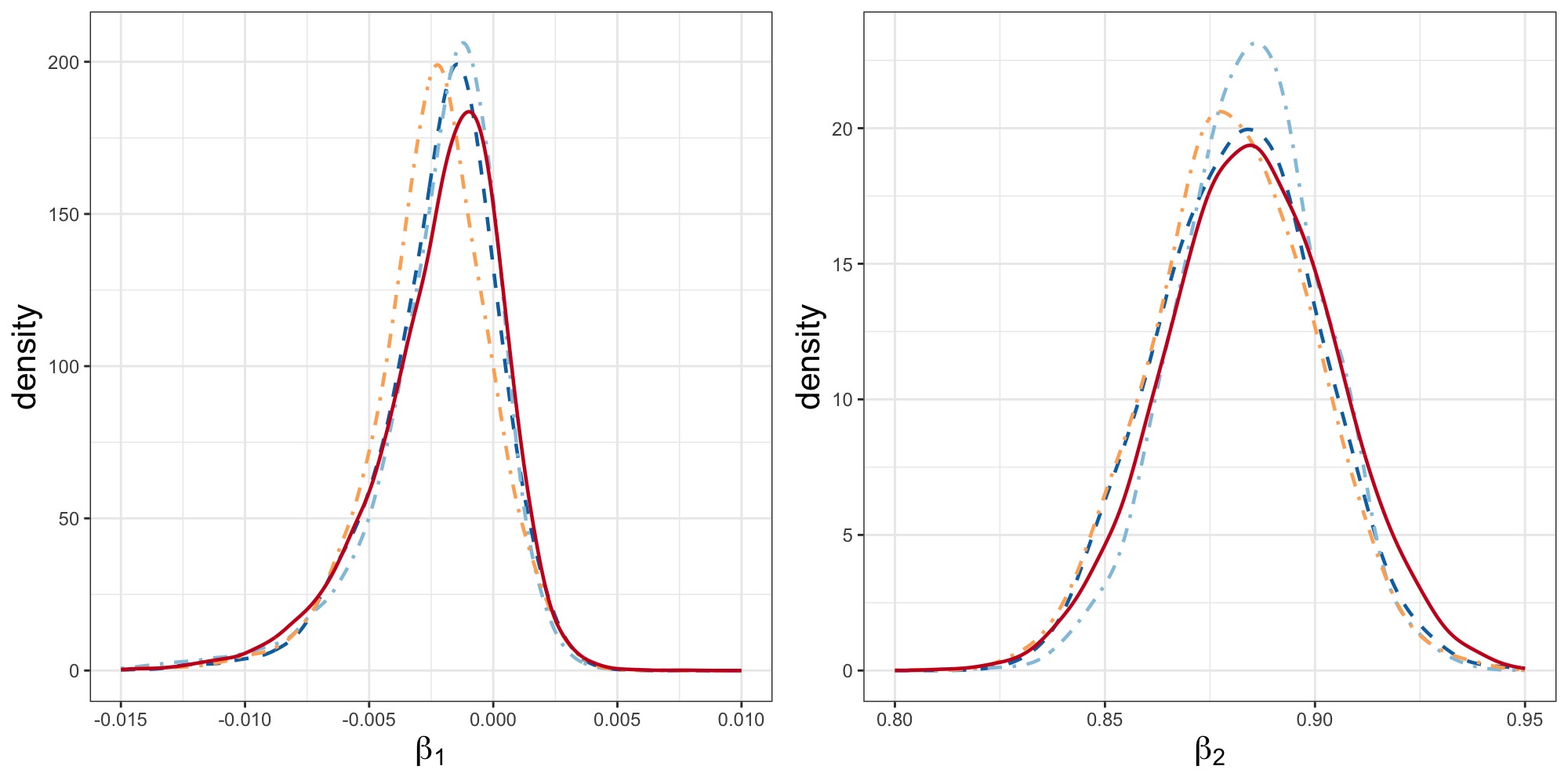}
        \caption{Marginal posteriors: red for EA, lightblue for Boland, darkblue for eq. grid with cubic interpolation, orange for gradient-based}
    \end{subfigure}
    \caption{Ice floe example}
    \label{fig:IceFloe}
\end{figure}

We illustrate the two-label autologistic model from section \ref{sec:autologistic} with a practical example. Specifically, we investigate spatial dependence within a $40\times40$ satellite image of ice floes available in the \texttt{R} package \texttt{PAWL} \citep{PAWLR}. The dataset is displayed in Figure \ref{fig:IceFloeData}.\\

\begin{figure}
    \centering
    \includegraphics[width=0.95\linewidth]{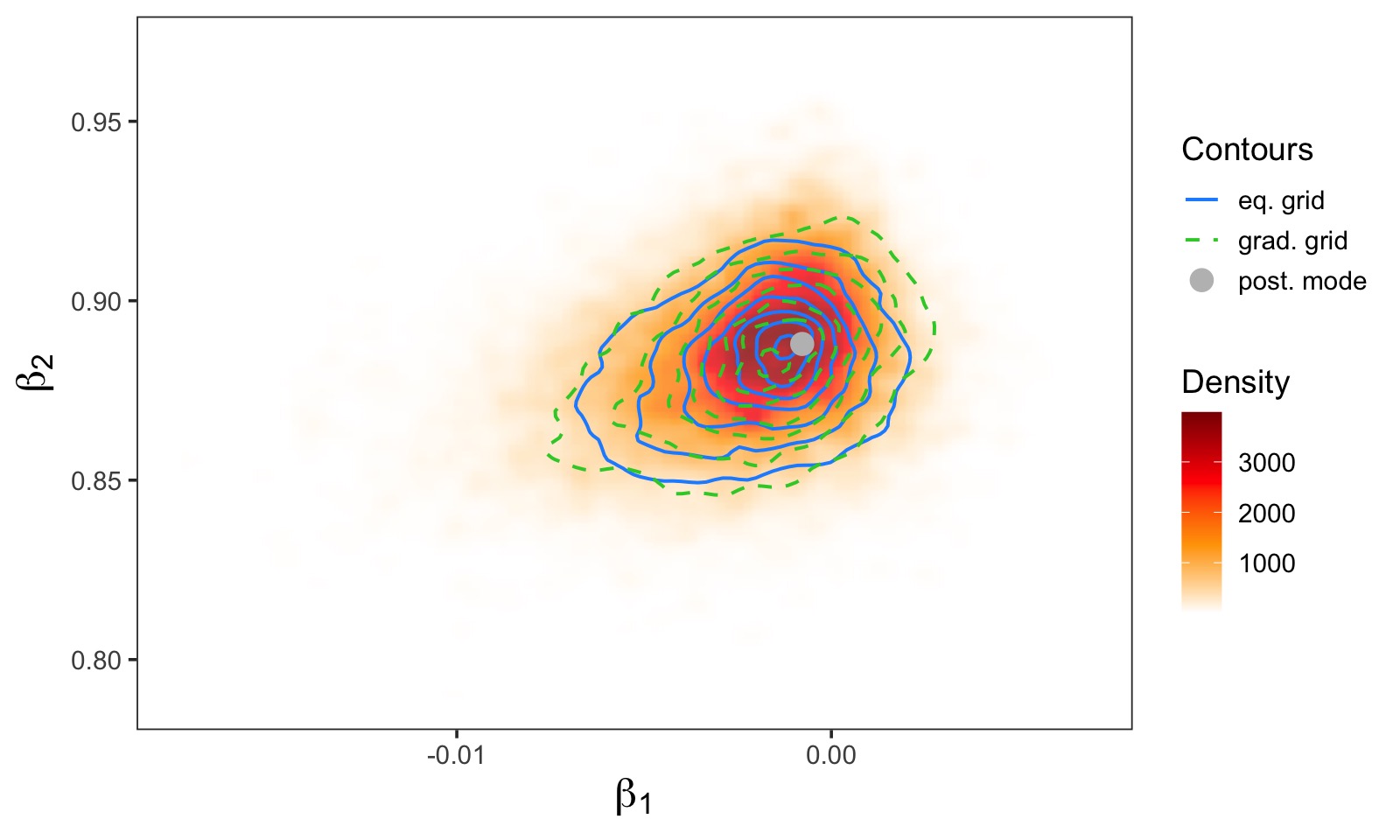}
    \caption{Volcano plot of the AEA posterior estimate with contours of  posteriors obtained with an eq. grid  and gradient-based grid in the autologistic example}
    \label{fig:2dposterior}
\end{figure}

Since model specific information is incorporated in our approach, we make slightly different choices than for the hidden Potts model. We keep boundaries for our grid, here the intervals $[-0.05, 0.05]$ for $\beta_1$ and $[0.5, 1.2]$ for $\beta_2$ as a stopping criterion. The equidistant grid on this parameter space has direction vectors $d_1=(0.003,0)^T$, $d_2=(0,0.05)^T$ and $n_{\beta}=1$, with a total grid size of $169$ points. Directional vectors and step size for the gradient-based approach are selected as described in section \ref{sec:pathchoices} with suitable chosen parameter $\kappa$. We propose $\beta_{crit} = (0,\log{(1+\sqrt{2})})$ as a starting point. Precomputation took $11.23$ minutes for the gradient based grid and $29.24$ minutes for the equidistant grid. The resulting grids are displayed in Figure \ref{fig:2dgrid}. Additionally to the exchange algorithm, we also compare to one of the methods described in \citet{boland2018gibbs}, whose strategy to compute one direction per dimension we also implemented. Estimates of the posterior can be found in Figure \ref{fig:2dposterior}. Computation time for theses estimates without precomputation was $5.7$ hours for the AEA, $0.1$ minutes for Boland, $0.5$ minutes for the equidistant grid and $0.02$ minutes for the gradient-based approach. Estimates of the marginal posterior distributions are shown in Figure \ref{fig:IceFloe}. The KL divergence to the posterior obtained through the AEA was $0.11$  in the equidistant grid approach, $0.15$ in the gradient-based grid approach and $0.15$ in the Boland approach.

\section{Conclusion}\label{sec:conclusion}

In this paper, we proposed a new approach for constructing sparse gradient-based grids over the space of sufficient statistics for Markov random field models with intractable normalizing constants. Through a simulation study, we established the accuracy of Hermite interpolation on these grids, enabling their use in the computation of Metropolis–Hastings acceptance ratios within MCMC algorithms. The resulting posterior estimates were evaluated in applications to satellite image classification using a hidden Potts model and to the analysis of spacial dependence in Arctic ice floes using an Ising model. Across these examples, our method achieved accuracy comparable to state-of-the-art methods such as the AEA and path sampling, while substantially reducing computational cost. \\

A key advantage of our proposed approach is its broad applicability and reliance on minimal model-specific assumptions. Unlike several existing approaches, our method naturally extends to models with higher-dimensional parameter spaces. This flexibility makes it a promising candidate for a wider range of applications, including social network modeling via exponential random graphical models \citep{frank1986markov} and the analysis of survey data using Ising models \citep{avalos2025bayesian}. At the same time, the practical feasibility of the approach in ultra-dimensional settings remains an important topic for further investigation. \\

Several avenues for future research emerge from this work. In particular, the selection of starting points for the grids currently relies on model-specific expertise. Incorporating closed-form biased posterior estimators, such as those proposed in \citet{laplante2025conjugategeneralisedbayesianinference}, may provide a principled and automated means of initializing grids. These approximations could also be combined with sparse-grid constructions to offer finer control over grid resolution and computational cost. Exploring such extensions represents a promising direction for further improving the scalability and robustness of grid-based inference for doubly-intractable models.

\section*{Acknowledgements}

The work of Laura Bazahica, Matthew Moores and Lassi Roininen was supported by the Research Council of Finland (Flagship of Advanced Mathematics for Sensing, Imaging and Modelling Grant 359183).

\bibliographystyle{apalike}
\bibliography{literature}

@article{moores2020scalable,
author = {Matthew T Moores and Geoff K Nicholls and Anthony N Pettitt and Kerrie L Mengersen},
title = {Scalable {B}ayesian Inference for the Inverse Temperature of a Hidden {P}otts Model},
volume = {15},
journal = {Bayesian Analysis},
number = {1},
publisher = {International Society for Bayesian Analysis},
pages = {1 -- 27},
year = {2020},
doi = {10.1214/18-BA1130},
}

@article{potts1952some,
  title={Some generalized order-disorder transformations},
  author={Potts, Renfrey B},
  journal={Mathematical Proceedings of the Cambridge Philosophical Society},
  volume={48},
number=1,
  pages={106--109},
  year={1952},
doi={10.1017/S0305004100027419}
}

@article{park2018bayesian,
  title={Bayesian inference in the presence of intractable normalizing functions},
  author={Park, Jaewoo and Haran, Murali},
  journal={Journal of the American Statistical Association},
  volume={113},
  number={523},
  pages={1372--1390},
  year={2018},
  publisher={Taylor \& Francis}
}

@article{gelman1998simulating,
  title={Simulating normalizing constants: From importance sampling to bridge sampling to path sampling},
  author={Gelman, Andrew and Meng, Xiao-Li},
  journal={Statistical science},
  volume={13},
  pages={163--185},
  year={1998},
  publisher={JSTOR}
}

@inproceedings{murray2012mcmc,
  title={{MCMC} for doubly-intractable distributions},
  author={Murray, Iain and Ghahramani, Zoubin and MacKay, David J C},
  booktitle={Proceedings of the 22nd Annual Conference on Uncertainty in Artificial Intelligence (UAI)},
  pages={359--366},
  year={2006},
  organization={AUAI Press}
}

@article{moller2006efficient,
  title={An efficient {M}arkov chain {M}onte {C}arlo method for distributions with intractable normalising constants},
  author={M{\o}ller, Jesper and Pettitt, Anthony N and Reeves, Robert and Berthelsen, Kasper K},
  journal={Biometrika},
  volume={93},
  number={2},
  pages={451--458},
  year={2006},
  publisher={Oxford University Press}
}

@article{cucala2009bayesian,
  title={A {B}ayesian reassessment of nearest-neighbor classification},
  author={Cucala, Lionel and Marin, Jean-Michel and Robert, Christian P and Titterington, D Michael},
  journal={Journal of the American Statistical Association},
  volume={104},
  number={485},
  pages={263--273},
  year={2009},
  publisher={Taylor \& Francis}
}

@article{boland2018gibbs,
author = {Aidan Boland and Nial Friel and Florian Maire},
title = {Efficient {MCMC} for {G}ibbs random fields using pre-computation},
volume = {12},
journal = {Electronic Journal of Statistics},
number = {2},
pages = {4138 -- 4179},
year = {2018},
doi = {10.1214/18-EJS1504},
}

@book{marin2007bayesian,
  title={Bayesian Core: a Practical Approach to Computational Bayesian Statistics},
  author={Marin, Jean-Michel and Robert, Christian P},
  year={2007},
  publisher={Springer}
}

@article{kechriniotis2024classical,
  title={Classical multivariate {H}ermite coordinate interpolation on $n$-dimensional grids},
  author={Kechriniotis, Aristides I and Delibasis, Konstantinos K and Oikonomou, Iro and Tsigaridas, Georgios N},
  journal={Journal of Computational and Applied Mathematics},
  volume={449},
  pages={115962},
  year={2024},
  publisher={Elsevier}
}

@article{white2016predicting,
  title={Predicting health programme participation: A gravity-based, hierarchical modelling approach},
  author={White, Nicole and Mengersen, Kerrie},
  journal={Journal of the Royal Statistical Society Series C: Applied Statistics},
  volume={65},
  number={1},
  pages={145--166},
  year={2016},
  doi={10.1111/rssc.12111}
}

@article{diggle1997regression,
  title={Regression modelling of disease risk in relation to point sources},
  author={Diggle, Peter and Morris, Sara and Elliott, Paul and Shaddick, Gavin},
  journal={Journal of the Royal Statistical Society Series A: Statistics in Society},
  volume={160},
  number={3},
  pages={491--505},
  year={1997},
  publisher={Oxford University Press}
}

@article{vu2025warped,
  title={Warped gradient-enhanced {G}aussian process surrogate models for exponential family likelihoods with intractable normalizing constants},
  author={Vu, Quan and Moores, Matthew T and Zammit-Mangion, Andrew},
  journal={Bayesian Analysis},
  volume={20},
  number={2},
  pages={435--459},
  year={2025},
  publisher={International Society for Bayesian Analysis}
}

@article{besag1974spatial,
  title={Spatial interaction and the statistical analysis of lattice systems},
  author={Besag, Julian},
  journal={Journal of the Royal Statistical Society: Series B (Methodological)},
  volume={36},
  number={2},
  pages={192--225},
  year={1974},
  publisher={Wiley Online Library}
}

@article{moores2020bayesian,
  title={Bayesian computation with intractable likelihoods},
  author={Moores, Matthew T and Pettitt, Anthony N and Mengersen, Kerrie L},
  journal={Case Studies in Applied Bayesian Data Science: CIRM Jean-Morlet Chair, Fall 2018},
  pages={137--151},
  year={2020},
  publisher={Springer}
}

@book{kolaczyk2014statistical,
  title={Statistical Analysis of Network Data with {R}},
  author={Kolaczyk, Eric D and Cs{\'a}rdi, G{\'a}bor},
  year={2014},
  publisher={Springer},
doi={https://doi.org/10.1007/978-3-030-44129-6}
}

@article{besag1991bayesian,
  title={Bayesian image restoration, with two applications in spatial statistics},
  author={Besag, Julian and York, Jeremy and Molli{\'e}, Annie},
  journal={Annals of the Institute of Statistical Mathematics},
  volume={43},
  number={1},
  pages={1--20},
  year={1991},
  publisher={Springer}
}

@book{winkler2012image,
  title={Image analysis, random fields and {M}arkov chain {M}onte {C}arlo methods: a mathematical introduction},
  author={Winkler, Gerhard},
  year={2003},
  publisher={Springer},
edition={$2^{nd}$}
}

@article{avalos2025bayesian,
  title={Bayesian inference of multiple {I}sing models for heterogeneous public opinion survey networks},
  author={Avalos-Pacheco, Alejandra and Lazzerini, Andrea and Lupparelli, Monia and Stingo, Francesco C},
  journal={Journal of the Royal Statistical Society Series C: Applied Statistics},
  pages={1395--1426},
  volume={74},
  year={2025},
  doi={https://doi.org/10.1093/jrsssc/qlaf028}
}

@article{frank1986markov,
  title={Markov graphs},
  author={Frank, Ove and Strauss, David},
  journal={Journal of the American Statistical Association},
  volume={81},
  number={395},
  pages={832--842},
  year={1986},
  doi={10.1080/01621459.1986.10478342}
}

@article{propp1996exact,
  title={Exact sampling with coupled {M}arkov chains and applications to statistical mechanics},
  author={Propp, James Gary and Wilson, David Bruce},
  journal={Random Structures \& Algorithms},
  volume={9},
  number={1-2},
  pages={223--252},
  year={1996},
  doi={10.1002/(SICI)1098-2418(199608/09)9:1/2<223::AID-RSA14>3.0.CO;2-O}
}

@book{huber2016perfect,
  title={Perfect Simulation},
  author={Huber, Mark L},
  year={2016},
publisher={Taylor \& Francis/CRC Press}
}

@article{calderhead2009estimating,
  title={Estimating {B}ayes factors via thermodynamic integration and population {MCMC}},
  author={Calderhead, Ben and Girolami, Mark},
  journal={Computational Statistics \& Data Analysis},
  volume={53},
  number={12},
  pages={4028--4045},
  year={2009},
  publisher={Elsevier}
}

@article{oates2016controlled,
  title={The controlled thermodynamic integral for {B}ayesian model evidence evaluation},
  author={Oates, Chris J and Papamarkou, Theodore and Girolami, Mark},
  journal={Journal of the American Statistical Association},
  volume={111},
  number={514},
  pages={634--645},
  year={2016},
  publisher={Taylor \& Francis}
}

@article{geman1984stochastic,
  title={Stochastic relaxation, {G}ibbs distributions, and the {B}ayesian restoration of images},
  author={Geman, Stuart and Geman, Donald},
  journal={IEEE Transactions on Pattern Analysis and Machine Intelligence},
  volume={PAMI-6},
  number={6},
  pages={721--741},
  year={1984},
  doi={10.1109/TPAMI.1984.4767596}
}

@article{swendsen1987nonuniversal,
  title={Nonuniversal critical dynamics in {M}onte {C}arlo simulations},
  author={Swendsen, Robert H and Wang, Jian-Sheng},
  journal={Physical Review Letters},
  volume={58},
  number={2},
  pages={86},
  year={1987},
  doi={10.1103/PhysRevLett.58.86}
}

@book{brooks2011handbook,
  title={Handbook of {M}arkov chain {M}onte {C}arlo},
  author={Brooks, Steve and Gelman, Andrew and Jones, Galin and Meng, Xiao-Li},
  year={2011},
  publisher={CRC press}
}

@article{robbins1951stochastic,
  title={A stochastic approximation method},
  author={Robbins, Herbert and Monro, Sutton},
  journal={The annals of mathematical statistics},
  volume={22},
  pages={400--407},
  year={1951},
  publisher={JSTOR}
}

@incollection{Avalos2026ABMI,
series = {Springer Proceedings in Mathematics \& Statistics},
issn = {2194-1009},
pages = {49--56},
publisher = {Springer Nature Switzerland},
booktitle = {New Trends in Bayesian Statistics},
isbn = {3031990080},
year = {2026},
title = {A {B}ayesian Multiple {I}sing Model},
copyright = {The Author(s), under exclusive license to Springer Nature Switzerland AG 2026},
language = {eng},
address = {Cham},
author = {Avalos-Pacheco, Alejandra and Lazzerini, Andrea and Lupparelli, Monia and Stingo, Francesco Claudio},
}

@misc{laplante2025conjugategeneralisedbayesianinference,
      title={Conjugate Generalised {B}ayesian Inference for Discrete Doubly Intractable Problems}, 
      author={William Laplante and Matias Altamirano and Jeremias Knoblauch and Andrew Duncan and François-Xavier Briol},
      year={2025},
      eprint={2511.23275},
      archivePrefix={arXiv},
      primaryClass={stat.ME},
      url={https://arxiv.org/abs/2511.23275}, 
}

@article{augustin1996autologistic,
  title={An autologistic model for the spatial distribution of wildlife},
  author={Augustin, NH and Mugglestone, Moira A and Buckland, Stephen T},
  journal={Journal of Applied Ecology},
  volume={33},
  pages={339--347},
  year={1996},
  publisher={JSTOR}
}

@article{wang2023recent,
  title={Recent advances in {B}ayesian optimization},
  author={Wang, Xilu and Jin, Yaochu and Schmitt, Sebastian and Olhofer, Markus},
  journal={ACM Computing Surveys},
  volume={55},
  number={13s},
  pages={1--36},
  year={2023},
  publisher={ACM New York, NY}
}

@article{jones1998efficient,
  title={Efficient global optimization of expensive black-box functions},
  author={Jones, Donald R and Schonlau, Matthias and Welch, William J},
  journal={Journal of Global optimization},
  volume={13},
  number={4},
  pages={455--492},
  year={1998},
  publisher={Springer}
}

@article{kullback1951information,
  title={On information and sufficiency},
  author={Kullback, Solomon and Leibler, Richard A},
  journal={The annals of mathematical statistics},
  volume={22},
  number={1},
  pages={79--86},
  year={1951},
  publisher={JSTOR}
}

@Manual{BayessR,
    title = {bayess: Bayesian Essentials with R},
    author = {Jean-Michel Marin and Christian P. Robert},
    year = {2024},
    note = {R package version 1.6},
    url = {https://CRAN.R-project.org/package=bayess},
    doi = {10.32614/CRAN.package.bayess},
  }

@Manual{PAWLR,
    title = {PAWL: Implementation of the PAWL algorithm},
    author = {Luke Bornn and Pierre E. Jacob},
    year = {2012},
    note = {R package version 0.5},
    url = {https://CRAN.R-project.org/package=PAWL},
    doi = {10.32614/CRAN.package.PAWL},
  }

@Manual{R,
    title = {R: A Language and Environment for Statistical Computing},
    author = {{R Core Team}},
    organization = {R Foundation for Statistical Computing},
    address = {Vienna, Austria},
    year = {2025},
    url = {https://www.R-project.org/},
  }

@article{moores2015external,
  title={An external field prior for the hidden Potts model with application to cone-beam computed tomography},
  author={Moores, Matthew T and Hargrave, Catriona E and Deegan, Timothy and Poulsen, Michael and Harden, Fiona and Mengersen, Kerrie},
  journal={Computational Statistics \& Data Analysis},
  volume={86},
  pages={27--41},
  year={2015},
  publisher={Elsevier}
}

@article{lux2021interpolation,
  title={Interpolation of sparse high-dimensional data},
  author={Lux, Thomas CH and Watson, Layne T and Chang, Tyler H and Hong, Yili and Cameron, Kirk},
  journal={Numerical Algorithms},
  volume={88},
  number={1},
  pages={281--313},
  year={2021},
  publisher={Springer}
}

\end{document}